\documentclass{aa}
\usepackage{graphics,epsfig,psfig,natbib}

\def\chandra{{\it Chandra}}

\def\hst{{\it HST}}

\def\rosat{{\it ROSAT}}

\def\lum{erg s$^{-1}$}
\def\flux{erg cm$^{-2}$ s$^{-1}$}

\def\arcsec{$^{\prime\prime}$}
\def\arcmin{$^{\prime}$}

\def\ltsima{$\; \buildrel < \over \sim \;$}
\def\simlt{\lower.5ex\hbox{\ltsima}} 
\def\gtsima{$\; \buildrel > \over \sim \;$}
\def\simgt{\lower.5ex\hbox{\gtsima}}

\begin{document}

\title{{\it Chandra} Observations of the X--ray Environment of BL~Lacs}

\author{D. Donato\inst{1} \and M. Gliozzi\inst{2} \and 
R. M. Sambruna\inst{1,2} \and J. E. Pesce\inst{2,3}}

\institute{George Mason University, School of Computational Sciences, 
4400 University Drive, Fairfax, VA 22030
\and George Mason University, Dept. of Physics \& Astronomy, MS 3F3,
4400 University Drive, Fairfax, VA 22030
\and Eureka Scientific}

\offprints{davide@physics.gmu.edu} 

\date{Received: ; accepted: }

\abstract{We present {\it Chandra} observations of the X--ray
environment of a sample of 6 BL Lacertae objects. The improved
sensitivity of the ACIS experiment allows us to separate the core
X-ray emission from the contribution of diffuse emission from the host
galaxy/cluster scales. 
Within the short (2--6 ks) ACIS exposures, we find evidence for
diffuse X--ray emission in 3 sources (BL Lac, PKS 0548--322, and PKS
2005--489). The diffuse emission can be modeled with a King profile
with $\beta \sim 0.3-0.6$, core radii $r_{\rm c} \sim 15 - 28$ kpc, and
0.4--5 keV luminosities in the range $10^{41}-10^{42}$ erg
s$^{-1}$. In the remaining 3 sources, one (3C 371) has a radial
profile entirely consistent with an unresolved source, while two (1ES
2344+514 and 1ES 2321+419) show evidence for weak diffuse emission on
kpc scales. These results support current models for radio-loud AGN
unifying BL Lacs and FRI radio galaxies through the orientation of
their jets. In PKS~0548--322 and PKS~2005--489, we also find evidence
for diffuse emission on cluster scales, although the spatial
properties of this emission are not constrained.  The temperature ($kT
\sim 3-5$ keV) and luminosity (L$_{0.4-5 \rm keV} \sim 10^{42}$ \lum)
of the cluster gas are typical of normal clusters. Interestingly,
these are the two brightest sources of the sample, suggesting a link
between environment and nuclear activity. 
}

\titlerunning{{\it Chandra} Observations of the X--ray Environment of BL~Lacs}
\authorrunning{D. Donato et al.\ }

\maketitle

\section{Introduction}
  
Orientation-based unification models for Active Galactic Nuclei have
been successful in explaining the rich variety of observed properties
in the various classes of AGN (Urry \& Padovani 1995; Antonucci
1993). According to these schemes, the various AGNs are the
same intrinsic object, powered by accretion of the host galaxy gas
onto a super-massive black hole, seen at different orientation angles
with respect to a preferred axis. In the case of radio-loud AGN, the
different subclasses are due to the different orientation of the
relativistic jet, with blazars (BL Lacertae objects and Flat Spectrum
Radio Quasars) corresponding to the more aligned sources, and radio
galaxies (Fanaroff-Riley I and II) being their parent populations
(e.g., Urry \& Padovani 1995 and references therein).

Previous studies at radio, IR, and optical wavelengths show that for
BL~Lacs (the relatively local, low-luminosity version of blazars) the
parent population is most likely represented by FR I radio galaxies,
although some sources may be occasionally hosted by FR IIs (e.g.,
Kollgaard et al. 1992). Indeed, it is possible to predict the
beamed luminosity function of BL~Lacs from the luminosity function of
FRIs invoking a Lorentz factor $\gamma \sim3-20$, depending on the
wavelength (Padovani \& Urry 1991; Urry et al. 1991).

A model-independent probe of these ideas is represented by studies of
the near environment. If FRIs are misaligned versions of BL~Lacs, the
larger-scale environment (host galaxy and cluster of galaxies) of the
two classes should be the same, as they are not affected by beaming.
Studies at optical-IR wavelengths showed that BL~Lacs and FRIs reside
in giant elliptical galaxies and, on average, in poor clusters of Abell
richness class 0 or less (Falomo et al. 1999, Wurtz et al. 1997). 

Similarly, the X-ray environment of BL~Lacs and FRIs should be
similar. {\it ROSAT}, and more recently {\it Chandra}, observations of
FRIs established that the X-ray cores of these sources are usually
embedded in diffuse soft X-ray emission which is interpreted as the
thermal halo of the host galaxy on kpc scales (Worrall et al. 2001, 
Canosa et al. 1999). However, because of the limited
sensitivity and spatial resolution of {\it ROSAT}, it was not possible to
perform similar observations for most BL~Lacs, where the bright core
dominates the X-ray emission. The only exception is PKS~0521--365,
where an unusually large and bright X-ray halo was detected with the
\rosat\ HRI (Hardcastle at al. 1999). 

Separating the diffuse X-ray emission from the core X-rays requires
high angular resolution and improved sensitivity, such as afforded by
the {\it Chandra} X-ray Observatory (e.g., Birkinshaw et al. 2002).  
Motivated by these considerations, we acquired
{\it Chandra} imaging observations of a sample of nearby ($z<0.1$)
BL~Lacs during cycle 1, to study their kpc-scale X-ray environment and
to compare to those of FRIs. As our sample contains sources claimed to
reside in optical clusters of galaxies (Falomo et al. 1999), {\it
Chandra} observations can be used to confirm independently the
presence of the cluster and quantify physical properties of the
intracluster gas.

Based on their spectral
energy distributions, BL~Lacs can be classified as Low-energy peaked
BL~Lacs (LBLs), when the radio-to-X-ray spectral index ($\alpha_{\rm rx}$)
is larger than 0.75; and as High-energy peaked BL~Lacs (HBLs), when
$\alpha_{\rm rx}$\ltsima0.75 (Padovani \& Giommi 1995). There is much
debate on the origin of the LBL-HBL division, which appears to be
continuous with luminosity (Fossati et al. 1997; Sambruna et
al. 1996). One of the goals of our GO1 \chandra\ proposal was to
investigate whether LBLs and HBLs exist in different X-ray
environments, to ascertain whether the ambient gas can affect their
different observed jet properties.

The outline of the paper is as follows. In Sect. 2 we discuss the sample
selection criteria and in Sect. 3 the observations and data analysis.
Results of the spatial and spectral are given in Sects. 4 and 5, respectively. 
The conclusions are presented in Sect. 6. The analysis of serendipitous X-ray
sources is given in Appendix A. Throughout this
paper, $H_0=75$ km s$^{-1}$ Mpc$^{-1}$ and $q_0=0.5$ are adopted.
With this choice, 1\arcmin\ corresponds to 71 kpc for PKS~0548--322
and for BL~Lac, 54 kpc for 3C 371, 73 kpc for PKS 2005--489, 62 kpc for 
1ES 2321+419, and 47 kpc for 1ES 2344+514.

\section{Sample selection} 

The targets were selected from two complete, flux-limited samples: the
1 Jy sample of radio-selected BL Lacs, and the {\it Einstein} Slew
Survey (1ES) sample of X-ray selected BL Lacs. Sources with redshifts 
larger than $z$=0.2 were excluded, to optimize
the {\it Chandra} resolution and for comparison with available samples
of FRIs (Worrall et al. 2001). Sources with $z<0.03$ were also
excluded, because the physical size of the field of view is too small
to study the intracluster gas ($\leq0.5$ Mpc), should they reside in
a cluster of galaxies. In this range we found 4 LBLs in the 1 Jy
sample. For each one we chose 2 HBLs matched in redshift (HBLs are
more common at low redshifts), in order to have similar redshift
distributions for the two types of BL~Lacs. The final list contained
12 targets. Only 6 targets were awarded {\it Chandra} time, including
4 HBLs and 2 LBLs. The targets are listed in Table 1, together with
their basic properties and classification.

Imaging data at longer wavelengths are available for all 6 sources,
for comparison with the ACIS images. Two of the sources (PKS 0548--322
and PKS 2005--489) are known to reside in a relatively rich optical
cluster of galaxies (Falomo et al. 1995; Pesce et al. 1995). The 
{\it Chandra} observation of 3C~371, where an X-ray
counterpart to the optical jet was found, was previously discussed in
Pesce et al. (2001), with main emphasis on the jet properties. Here,
we will neglect the X-ray jet for this source and focus on the
extended X-ray environment. 


\begin{table*}
\caption{Targets: Basic Properties}
\begin{center}
\begin{tabular}{lccccccc}
\hline
\noalign{\smallskip}
Object Name  & R.A.        &  Dec.        & z & $N_{\rm H,Gal}$ & Type & $L_{2-10 {\rm keV}}$&Ref.\\
~~~~~~~(1)    & (2)         &  (3)         &(4)&     (5)         & (6)  &     (7)             &(8) \\
\noalign{\smallskip}
\hline
\noalign{\smallskip}
PKS 0548--322 & 05 50 40.80 & -32 16 17.80 & 0.069 & 2.52  & HBL & 31.9 & KU98\\ 
3C 371       & 18 06 50.60 & +69 49 28.10 & 0.051 & 4.73  & LBL &  1.5 & DO01\\  
PKS 2005--489 & 20 09 25.40 & -48 49 54.00 & 0.071 & 5.08  & HBL & 62.2 & PA01\\ 
BL Lac       & 22 02 43.30 & +42 16 39.80 & 0.069 & 21.6  & LBL &  6.1 & RA02\\  
1ES 2321+419 & 23 23 52.10 & +42 10 59.00 & 0.059 & 10.9  & HBL &  1.5 & PE96\\  
1ES 2344+514 & 23 47 04.80 & +51 42 17.40 & 0.044 & 16.8  & HBL &  3.9 & DO01\\  
\noalign{\smallskip}
\hline
\end{tabular}
\end{center}
{\bf Columns}: 1=Source name; 2=Right Ascension at J2000; 3=Declination at J2000; 4=Redshift; 5=Galactic column density in units of $10^{20} {\rm cm^{-2}}$. The Galactic values was derived from the nh program at
HEASARC (based on Dickey \& Lockman 1990); 6=Source type: HBL=High--Energy Peaked BL Lacertae object, LBL=Low--Energy Peaked BL Lacertae object; 7=Intrinsic X--ray luminosity (in units of $10^{43}$ erg s$^{-1}$) between 2 and 10 keV from data in literature; 8=Reference for the X--ray luminosity: KU98=Kubo et al. 1998; DO01=Donato et al. 2001; PA01=Padovani et al. 2001; RA02=Ravasio et al. 2002; PE96=Perlman et al. 1996.
\end{table*}

\section{Observations and Data Reduction}

The {\it Chandra} observations of the 6 objects were carried out in
early 2000. A log of the \chandra\ observations is reported in
Table 2, together with information about the source intensities. All
sources were observed with ACIS-I at the aimpoint of the I3 chip,
except 3C~371 which was observed with ACIS-S, at the aimpoint of S3.
The choice of S3 for 3C~371 is due to the fact that it was known to
have a synchrotron optical jet.  As a science goal for this target was
to find the X-ray counterpart of the synchrotron jet, the softer
response of a back-illuminated CCD such as S3
was needed to maximize the detection of the X-ray jet. 

The short exposures (2--6 ks) were designed to search for the X-ray
diffuse emission around the cores, but are clearly insufficient to
study in detail their physical properties. Again an exception is
3C~371, where an exposure of 10 ks was requested, in order to detect
the X-ray jet (see, e.g., Sambruna et al. 2002). 

Since the main purpose of this work is to observe the large-scale
environments of blazars, the observations were made with the full CCD
operational and the standard ACIS frame time (3.24 sec), even if it
was expected that the central sources would have produced a high
X-ray count rate ($\geq1$ c/s). With this set-up it is likely that
more than one photon will be detected in a single CCD pixel within a
single integration time. This effect, called pile-up, leads to a
distortion (hardening) of the spectrum of the core and a 
reduction in the measured count rate relative to the true incoming 
rate. Pile-up becomes important for $\geq0.1$ c/frame. 

The pile-up effect also produces an image of the core that is
different from the original source image: it appears somewhat extended
and flat-topped, and in the cases of most extreme pile-up with a
central hole, where the pixels have no counts. In this case the
pile-up is strong enough that the total amplitude of the event is
larger than the on-board threshold ($\sim$15 keV) and is
rejected. This rejection is visible in the images of the central core
region of PKS 0548--322 and PKS 2005--489.

For the image analysis we used the \verb+CIAO+ 2.2.1 software package
and followed the standard reduction criteria, using the latest files
provided by the {\it Chandra} X--ray Center. We inspected the light
curves of the background in all cases to search for possible
background fluctuations. Only for PKS 0548--322 we found a short
variation of the background intensity. The corresponding time interval
was removed from the analysis. We restricted our analysis to the
0.3--8.0 keV energy range, where the instrument is better calibrated
and the background is negligible.

The exposure for BL~Lac was divided into two segments taken at
different times for operational reasons (Table 2). The 2 images of
BL~Lac were combined using the \verb+CIAO+ tool {\tt mergeall} in
order to increase the signal-to-noise ratio, for a total exposure of
3.1~ks. 


\begin{table*}
\caption{Targets: Observation Log and Data}
\begin{center}
\begin{tabular}{lccccccc}
\hline
\noalign{\smallskip}
Object Name  & Seq. Num.& Exposure & Start Date & Net Rate & Bkg Rate & c/f \\
~~~~~~~(1) & (2)      &  (3)     &(4)         &   (5)    &    (6)   & (7) \\
\noalign{\smallskip}
\hline
\noalign{\smallskip}
PKS 0548--322 & 700145 & 4559  &1999/12/31  &  0.235$\pm$0.007 & 4.71 & 0.7516 \\
3C 371        & 700146 & 10120 &2000/03/21  &  0.263$\pm$0.005 & 4.56 & 0.8334 \\ 
PKS 2005--489 & 700147 & 5858  &2000/10/07  &  0.491$\pm$0.009 & 2.61 & 1.5719 \\
BL Lac        & 700148 & 2117  &2000/01/07  &  0.171$\pm$0.009 & 0.14 & 0.5464 \\ 
              & 700196 & 1039  &2000/02/16  &  0.172$\pm$0.013 & 0.08 & ---    \\ 
1ES 2321+419  & 700149 & 4628  &2000/02/08  &  0.253$\pm$0.007 & 0.84 & 0.8114 \\ 
1ES 2344+514  & 700150 & 2727  &2000/02/08  &  0.185$\pm$0.008 & 1.17 & 0.5912 \\ 
\noalign{\smallskip}							      
\hline
\end{tabular}
\end{center}
{\bf Columns}: 1=Source name; 2=Sequence number of the observation; 3=Net Chandra exposure in seconds after data screening; 4=Observation start date (Year-Month-Day); 5=Net photon count rate (c/s) of the source in the energy range 0.3--8 keV; 6=Background photon count rate (in units of $10^{-3}$ s$^{-1}$) obtained rescaling to the same extraction region used for the net source count rate; 7=Count/frame evaluated with PIMMs and using spectral information from the literature (see previous Table for the references). The input flux in PIMMs is such that the output flux corresponds to the measured one. 
\end{table*}

\begin{figure*}
\noindent{\psfig{file=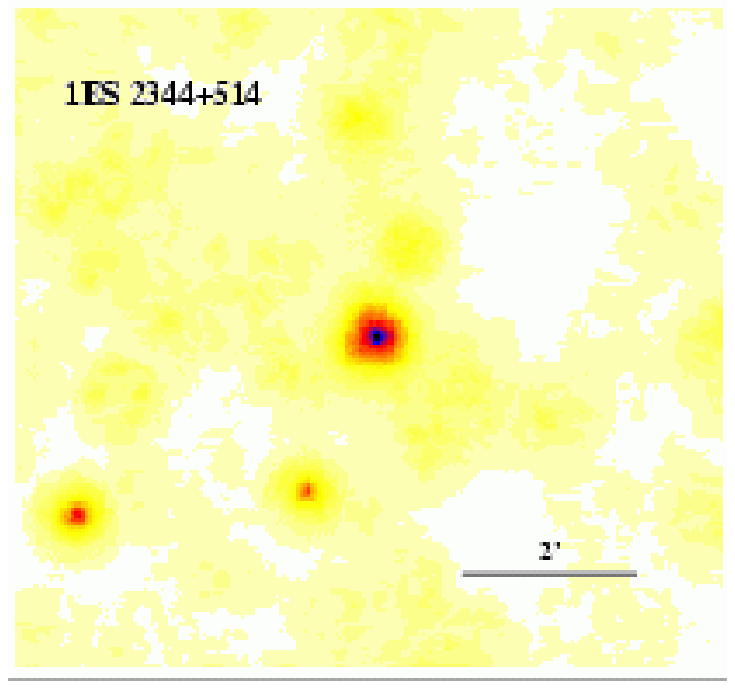,width=9cm,height=7.5cm}}{\psfig{file=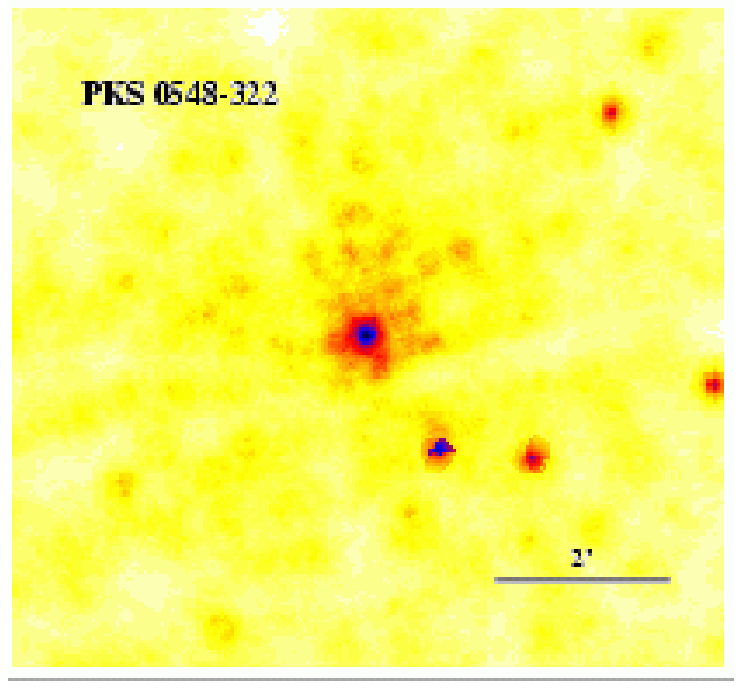,width=9cm,height=7.5cm}}
\caption[]{Adaptively-smoothed \chandra\ ACIS-I images of 2 BL Lacs.
North is up and East to the left. In PKS 0548--322, diffuse X-ray
emission on cluster scales is apparent, while the image of 1ES
2344+514 exhibits a point source embedded in diffuse emission on the
scale of the galaxy's halo.}
\end{figure*}

\section{Spatial Analysis}

The main aim of the spatial analysis is to investigate whether diffuse
emission is present around the BL Lac objects or whether their X-ray
radial profile is entirely consistent with a point source. In
the latter case, no contribution to the local X--ray emission from a
diffuse component is present.

The first step of the spatial analysis was to produce a smooth image
of the field of view.
For example, Fig. 1 shows the adaptively smoothed Chandra images 
of 1ES 2344+514 and PKS 0548-322, in the 0.3--8 keV energy range.

Before extracting the radial profiles of the sources, we removed
 from the images instrumental features (i.e., spikes produced by 
out-of-time events) and field sources. 
To find the field
sources we used the detect tool {\tt wavdetect} with the default value
($10^{-9}$) for the threshold for identifying a pixel as belonging to
a source (see Sect. 7). For 3C~371, we also removed the X-ray jet (Pesce
et al. 2001) by using an elliptical region (1.6\arcsec $\times$ 1.1\arcsec) 
centered on the jet. The same region was substituted by a
region of identical shape and size located symmetrically with respect
to the core, in order to restore a possible halo contribution.

Using the tools {\tt dmextract} and {\tt dmtcalc}, we extracted the
radial profiles on a scale typical of clusters (i.e.,
approximately 250 kpc, see for example De Grandi et al. 1999) that
correspond to a radius of about 200\arcsec~on our images, depending on
the redshifts. In the case of the ACIS-I observations, this radius
encompasses the boundary between the various CCDs. The estimated 
average influence of the gap between the single CCDs on the surface 
brightness is of the order of 4\%, therefore the surface brightness
should be considered as a lower limit. 
A series of annular regions, with increments of
the radius of 2\arcsec, was used to extract the radial profile. For
3C~371, the only source observed with ACIS-S, the number of the annuli
was limited by the distance between the source and the edge of S3.

While the responses of ACIS-I CCDs are not dramatically different,
the response of S3 and S2 are different. The background regions were
extracted from annular, circular, or box regions external to the
annular regions centered on the sources. These regions are free of
obvious sources.  

In order to evaluate the instrumental response to a point source, we
used the tool {\tt mkpsf} to create an image of the Point Spread
Function (PSF) of an on-axis point source, normalized to the source
flux. The PSF changes with source position and photon energy, and is
created by interpolation of the medium resolution library of
pre-launch calibration files (the PSF hypercube library). The
{\tt mkpsf} tool is able to create only monochromatic PSFs.
However, a monochromatic PSF is too simple a model to describe
adequately the spatial distribution of our sources.  We therefore
improved the model to find a better representation of the source PSF,
following a method similar to that used by Worrall et al. (2001).  Our
improvement consists on merging 8 different monochromatic PSFs chosen
and weighted on the basis of the source energy spectra between 0.3 and
8 keV. 

This method can be summarized as follows: 

1) We first extracted the energy distributions of the photons from a
circular region centered on the source. The radii of these regions
vary from 5\arcsec\ to 10\arcsec, depending on the source brightness.

2) In each case, we sampled the entire photon distribution by choosing
8 discrete energy values.  The number of counts at any of these energy
values corresponds to the energy weight (see Fig. 2).  The energies
and weights are shown in Table 3 and are indicated with E and W,
respectively. 

3) Using {\tt mkpsf} we created 8 monochromatic PSFs at the position
of the AGN on the detector for each of the 8 sampled energies, and
coadded them. Each PSF was weighted by its relative normalization (the
weights).

\begin{figure}[h]
\noindent{\psfig{file=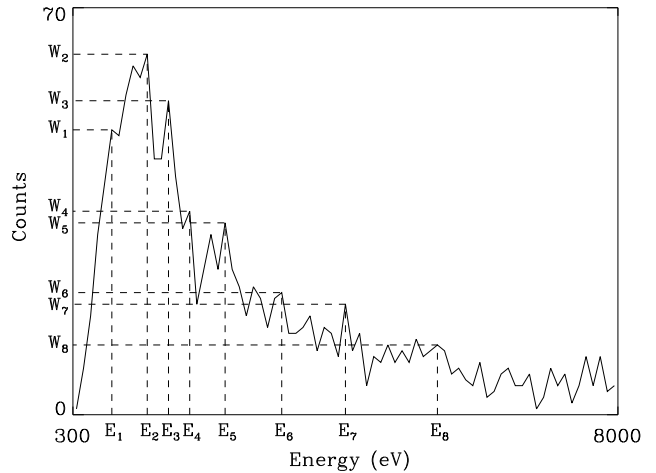,width=8.5cm,height=6.5cm,bbllx=53pt,bblly=64pt,bburx=576pt,bbury=468pt,clip=}}
\caption[]{Example of energy histogram for 1ES 2321+419, 
illustrating the procedure needed for the composite PSF (see text). 
The energies, E, and weights, W, for the monochromatic PSFs are
indicated by the dashed lines.}
\end{figure}


\begin{table*}
\caption{Energy weighting for the monochromatic PSFs}
\begin{center}
\begin{tabular}{lcccccccc}
\hline
\noalign{\smallskip}
Object Name  & $E_1, W_1$& $E_2, W_2$& $E_3, W_3$& $E_4, W_4$& $E_5, W_5$& $E_6, W_6$& $E_7, W_7$& $E_8, W_8$\\ 
~~~~~~~(1)    & (2)         &  (3)    &(4)        &     (5)   & (6)       &     (7)   &(8)        & (9)       \\
\noalign{\smallskip}
\hline
\noalign{\smallskip}
PKS 0548--322 &  0.82, 34 & 1.22,  38 & 1.37,  30 & 1.62, 21  & 1.87, 19  & 2.07, 11  & 2.77,  8  & 3.47,  7 \\
3C 371        &  0.47, 58 & 0.87,  58 & 1.17,  37 & 1.47, 37  & 1.87, 26  & 2.17, 23  & 2.72, 17  & 4.02, 10 \\
PKS 2005--489 &  0.57, 77 & 0.87, 117 & 1.17, 102 & 1.27, 91  & 1.42, 70  & 1.52, 63  & 1.77, 38  & 1.97, 36 \\
BL Lac        &  0.87,  7 & 1.22,  14 & 1.37,  10 & 1.62, 15  & 2.47,  8  & 3.07,  9  & 3.62,  6  & 4.17,  8 \\
1ES 2321+419  &  0.77, 21 & 1.12,  28 & 1.32,  32 & 1.62, 28  & 1.92, 17  & 2.47, 16  & 2.92, 13  & 3.67, 11 \\
1ES 2344+514  &  0.52,  5 & 1.02,  15 & 1.27,  15 & 1.52, 12  & 1.72,  7  & 2.17,  6  & 2.57,  6  & 4.82,  7 \\
\noalign{\smallskip}
\hline
\end{tabular}
\end{center}
{\bf Columns}: 1=Source name; 2-9=Energy (in keV) and weight of each monochromatic PSF that has been coadded in order to have a more realistic representation of the source PSF (see Sect. 4).

\end{table*}

\begin{table*}
\caption{PSF Parameter Values}
\begin{center}
\begin{tabular}{lcccccccc}
\hline
\noalign{\smallskip}
Object Name  & $A_0$& $A_1$    & $A_2$ & $A_3$    & $A_4$   & $A_5$& $A_6$   &$A_7$ \\
             &      & $\times 10^{-2}$&       & $\times 10^{-5}$&$\times 10^{-5}$&      &$\times 10^{-4}$&      \\
~~~~~~~(1)    & (2)  &  (3)     &(4)    &    (5)   & (6)     &  (7) & (8)     & (9)  \\
\noalign{\smallskip}
\hline
\noalign{\smallskip}
PKS 0548--322 & 0.17 & 5.5  &  5.2  &  0.5 & 0.7  & 0.34 & 1.4 & 12.5  \\
3C 371        & 0.17 & 3.0  &  9.0  &  1.0 & 1.1  & 0.36 & 1.0 & 16.6  \\
PKS 2005--489 & 0.29 & 6.5  & 10.0  &  3.7 & 4.8  & 0.09 & 2.0 & 12.0  \\
BL Lac        & 0.30 & 2.0  &  7.1  & 14.0 & 0.2  & 0.02 & 1.8 & 70.0  \\
1ES 2321+419  & 0.12 & 2.8  &  6.5  & 14.0 & 0.5  & 0.30 & 1.7 & 85.0  \\
1ES 2344+514  & 0.32 & 3.0  &  9.0  &  4.5 & 1.0  & 1.64 & 0.8 & 15.0  \\
\noalign{\smallskip}
\hline
\end{tabular}
\end{center}
{\bf Columns}: 1=Source name; 2-9=Values of the parameters of the analytical function used to fit the composed PSF (see Sect. 4 and Eq. (1)).
\end{table*}

Once we obtained the composite PSF, we searched for a fair analytical
representation that describes it. This empirical representation is
obtained using an 8-parameter function:

\begin{equation}
PSF(\rm{x})= A_0[A_1\frac{e^{-\frac{\rm{x}}{A_2}}}{\rm{x}^2}-A_3+A_4e^{-\frac{(\rm{x}-15)^2}{A_5}}+A_6e^{-\frac{(\rm{x}-13)^2}{A_7}}]
\end{equation}

For each source,
we fitted the composite PSF with Eq. (1) and found the best-fit
values of the 8 parameters (Table 4).  The instrumental radial profile
is not expected to be a good representation for the central region of
strongly piled-up sources. Therefore, since no \verb+CIAO+ task exists
yet to correct the PSF for pile-up, we normalized the
instrumental radial profile at $r$=2\arcsec:
at larger distances the pile-up effect on the PSF is
negligible, according to the {\it Chandra} Proposers' Observatory
Guide.  For each source we superimposed the composite PSF (weighted 
sum of the 8 monochromatic PSFs) on the surface brightness
versus the radial distance.

Due to the pile-up effect on the spectral distribution of the photons
(spectral hardening), this method is likely to overestimate the contribution
of the hard photons. The result is a slightly broader PSF, which will not 
cause any false detection of extended emission but rather underestimate 
its contribution.

After submitting this paper, a script that 
automatically performs the same task (\verb+ChaRT+)
was released by the \chandra\ X--ray Center (\verb+CXC+). We performed
the analysis of the radial profiles using \verb+ChaRT+ and found
consistent results with our method above.
In Fig. 3 we plot the radial profiles of the 6 sources. The dashed
line indicates the profile of the composite PSF obtained using Eq. (1)
and the parameters in Table 4. The horizontal dotted line represents
the background. 

In the case of 3C~371, the radial profile is consistent with the PSF
and no excess emission is detected in our exposure. In the case of BL
Lac, 1ES 2321+419, and 1ES 2344+514, there is some evidence for weak
excess emission between 18\arcsec\ and 30\arcsec.  Excess emission
over the PSF is clearly present for PKS 0548--322 and PKS 2005--489 up
to $\sim$ 150\arcsec\ (180 kpc) and 130\arcsec\ (160 kpc), respectively
(Fig. 3).

A $\beta$-model (King profile) was used to model the excess X-ray 
emission in all cases.
We find that the $\beta$ model provides a significant improvement only
in the case of BL Lac, PKS 0548--322, and PKS 2005--489.  The free
parameters of the $\beta$-model are the value of $\beta$ and the core
radius, $r_{\rm c}$.  For PKS 0548--322, the reduced $\chi^2$ of the fit is
$\chi^2_{\rm r}=1.23$ with 142 degrees of freedom (d.o.f.), with
$\beta=0.35 \pm$0.20 and $r_{\rm c}$=23.2\arcsec$\pm$2.5\arcsec.
For PKS 2005--489, $\chi^2_{\rm r}=1.0$ (138 d.o.f.), with $\beta=0.56
\pm$0.08 and $r_{\rm c}$=12.1\arcsec$\pm$1.2\arcsec.
For BL~Lac, the model parameters are poorly constrained; we thus 
fixed $\beta$ to 0.6 (Worrall \& Birkinshaw 1994) and left $r_{\rm c}$ as 
the only free parameter of the fit. We obtain $\chi^2_{\rm r}=0.94$ for 
24 d.o.f. and $r_{\rm c}$=14.7\arcsec$\pm$12.6\arcsec. 
All the errors are 1$\sigma$. 

The core radius implies a physical radius of the diffuse emission of
28 kpc in PKS 0548--322, 15 kpc in PKS 2005--489, and 18 kpc for
BL~Lac. These values are typical of the X-ray halo of the host galaxy
(Birkinshaw et al. 2002, Fabbiano et al. 1992), which is a giant
elliptical in all cases (Falomo \& Kotilainen 1999). Fig. 4 shows the PSF
plus $\beta$ model fit for PKS 0548--322 and PKS 2005--489, where the
parameters are best constrained.

As is apparent from the observed radial profiles of PKS 0548--322 and
PKS 2005--489, diffuse emission in these two sources extends well
beyond the fitted value of the core radii, up to $\sim$
130-150\arcsec, or several hundreds of kiloparsecs. This scale is
typical of clusters, and indeed both sources are known to reside in
optical clusters (Pesce et al. 1995). However, the radial profiles are
suggestive of two components, one (the galaxy halo) up to 40\arcsec,
plus a large-scale tail related to the cluster. 
Thus, we attempted a fit to the 

\clearpage
\begin{figure*}
\centering
{\includegraphics[width=9.1cm]{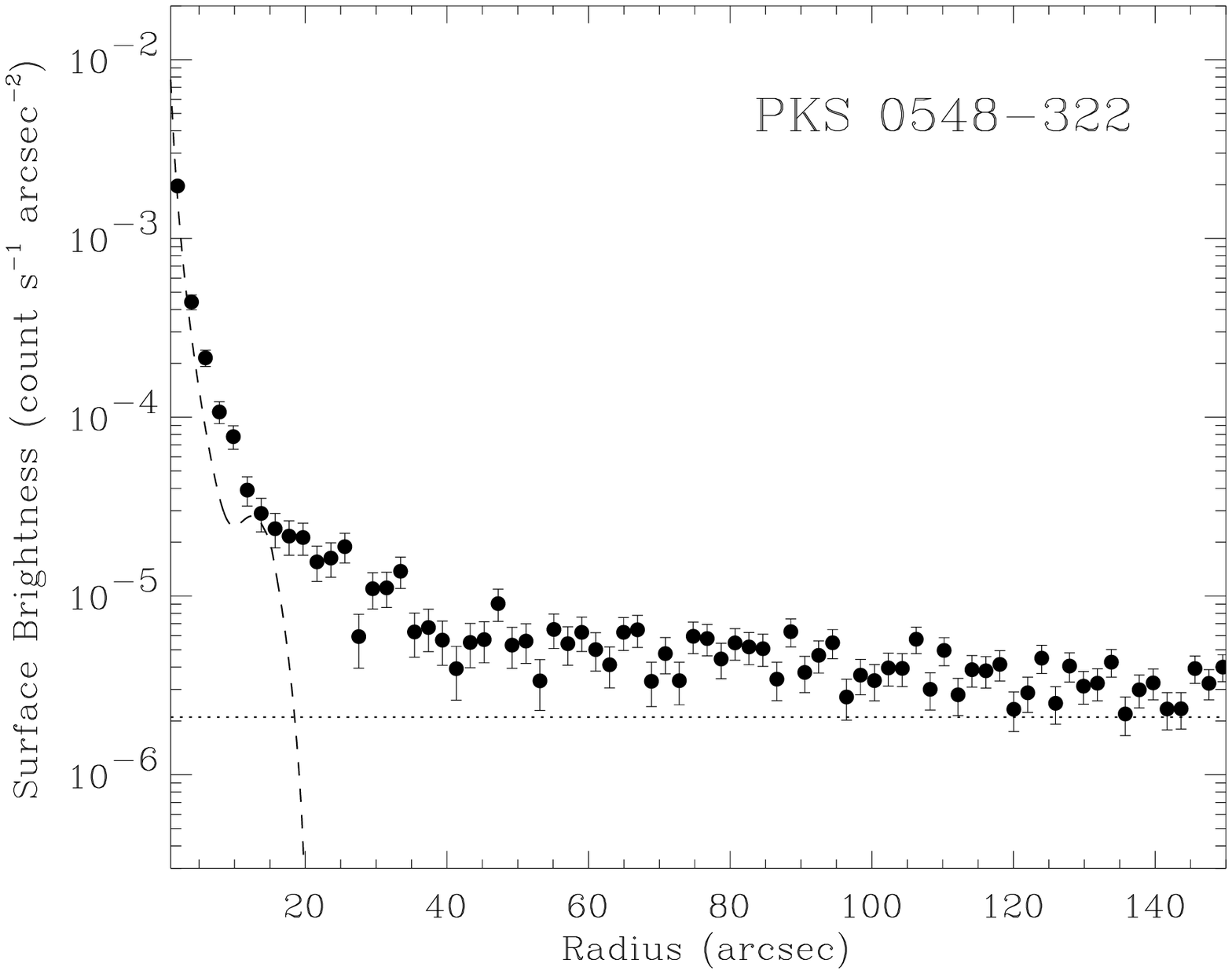}}{\includegraphics[width=9.1cm]{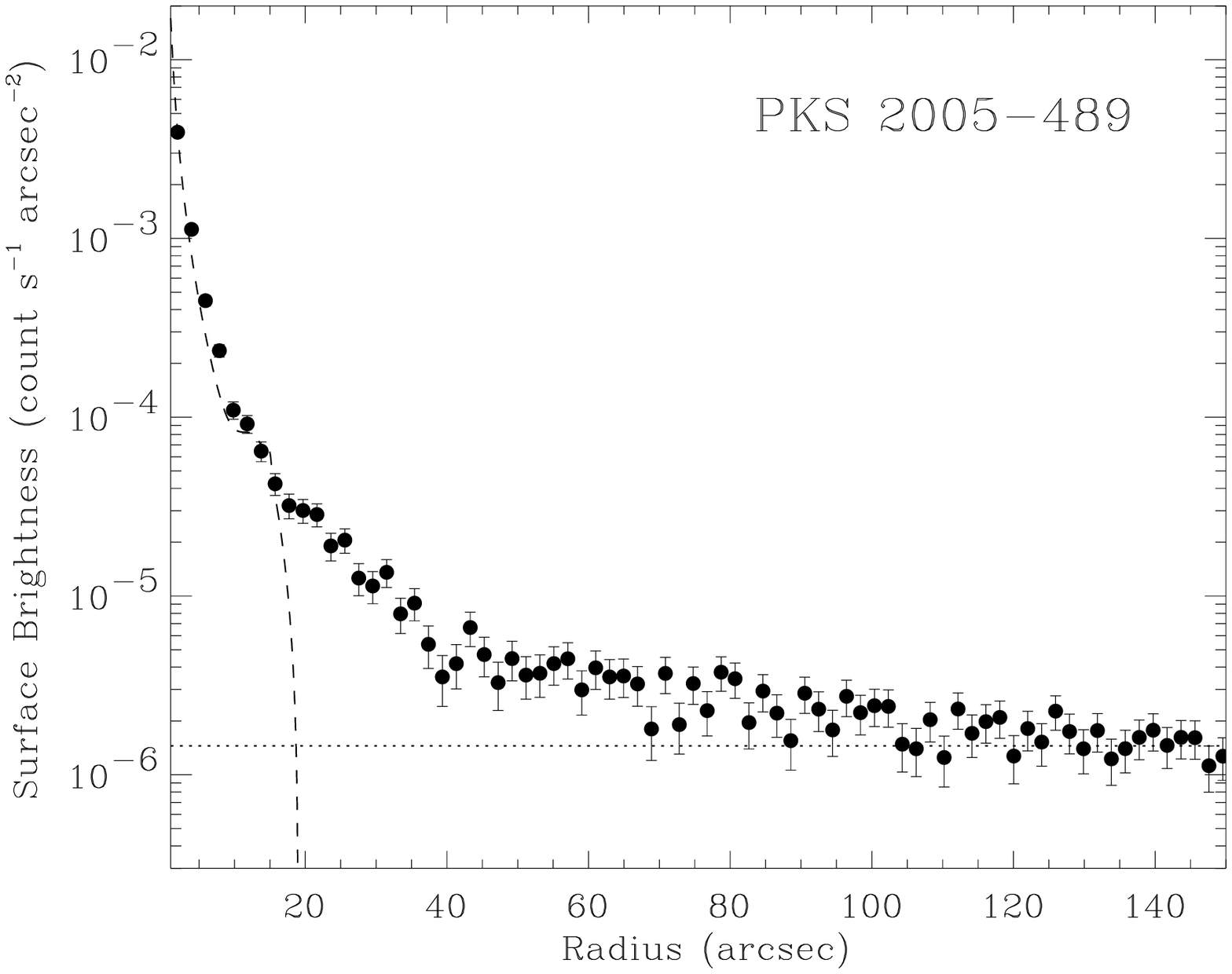}}
{\includegraphics[width=9.1cm]{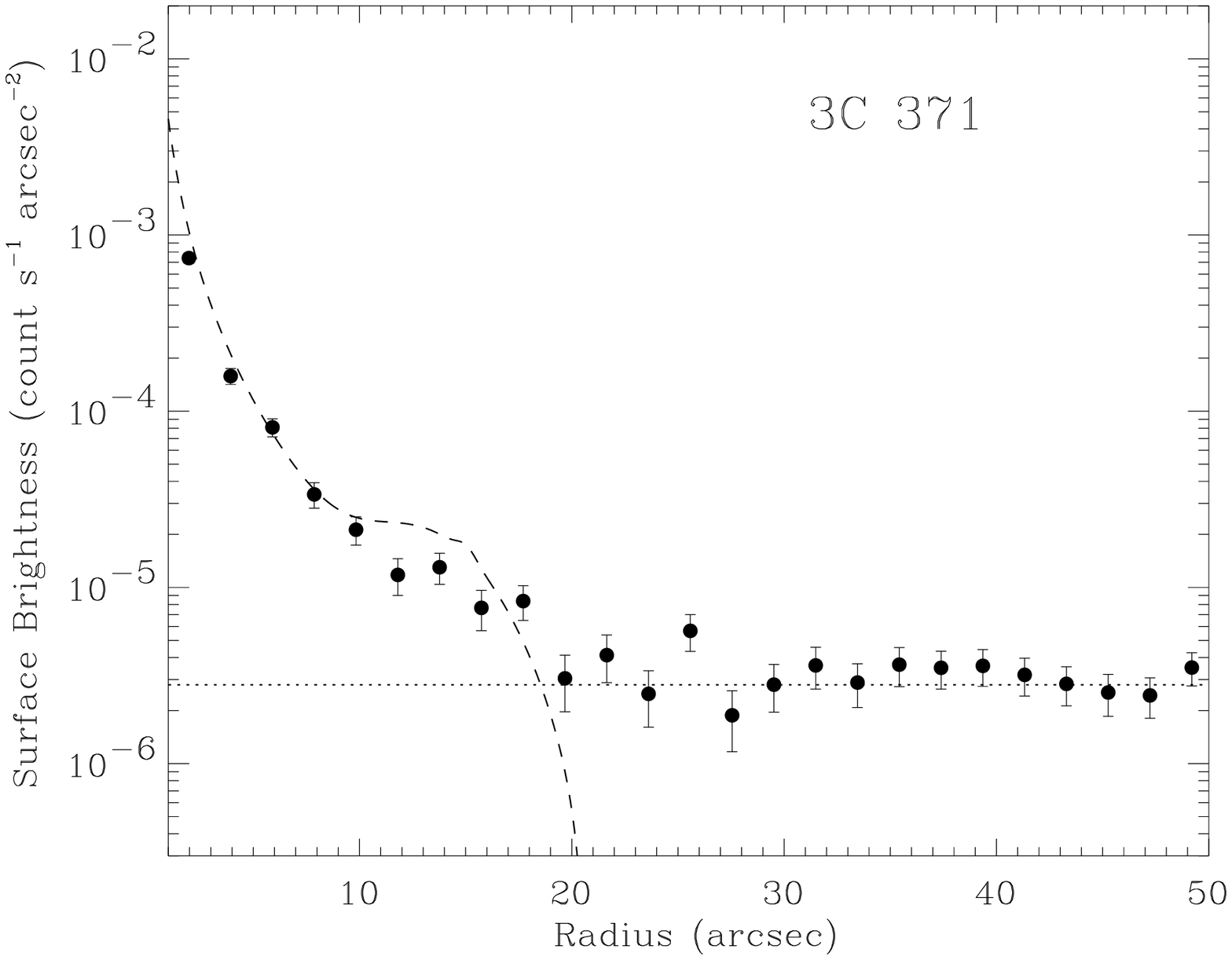}}{\includegraphics[width=9.1cm]{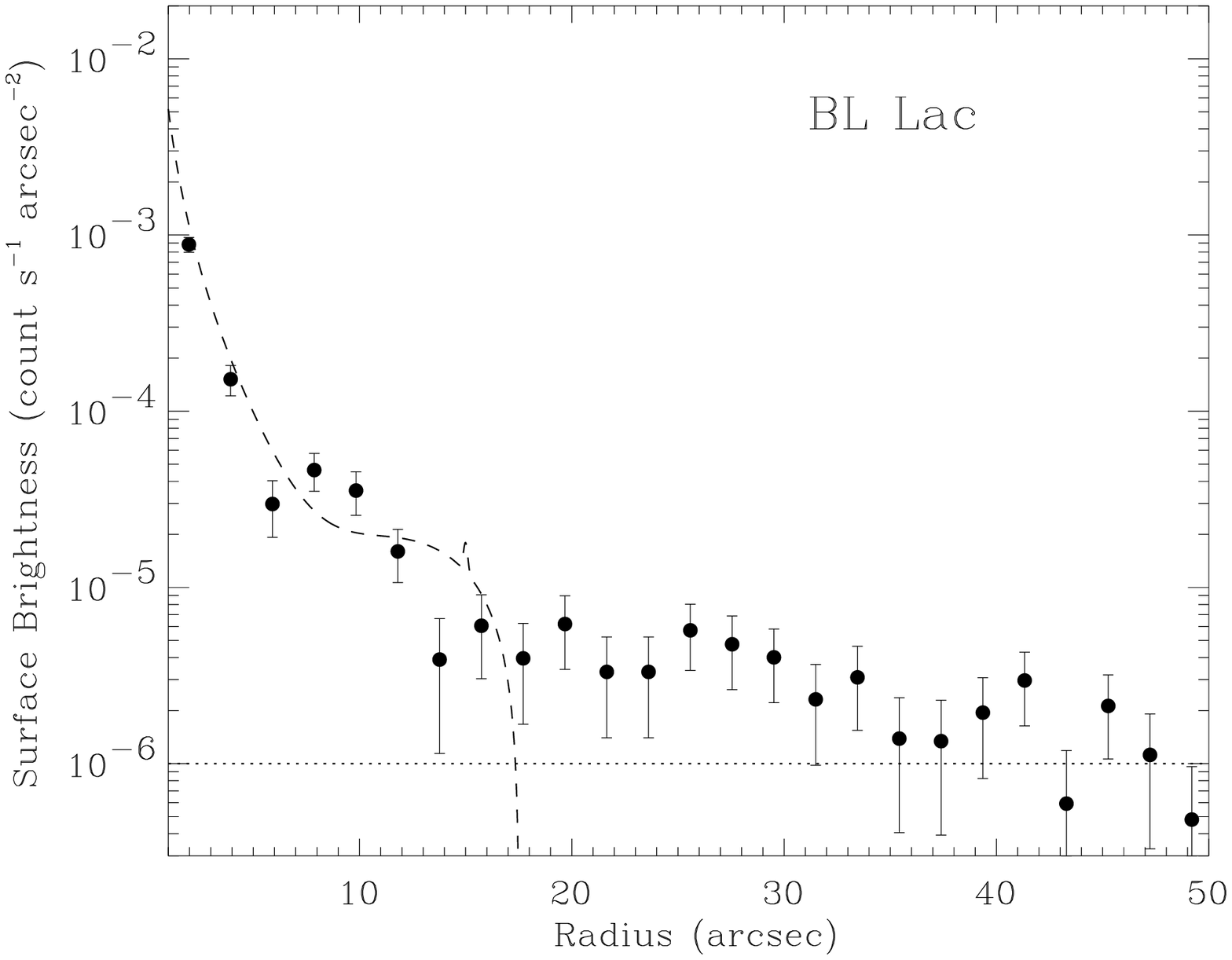}}
{\includegraphics[width=9.1cm]{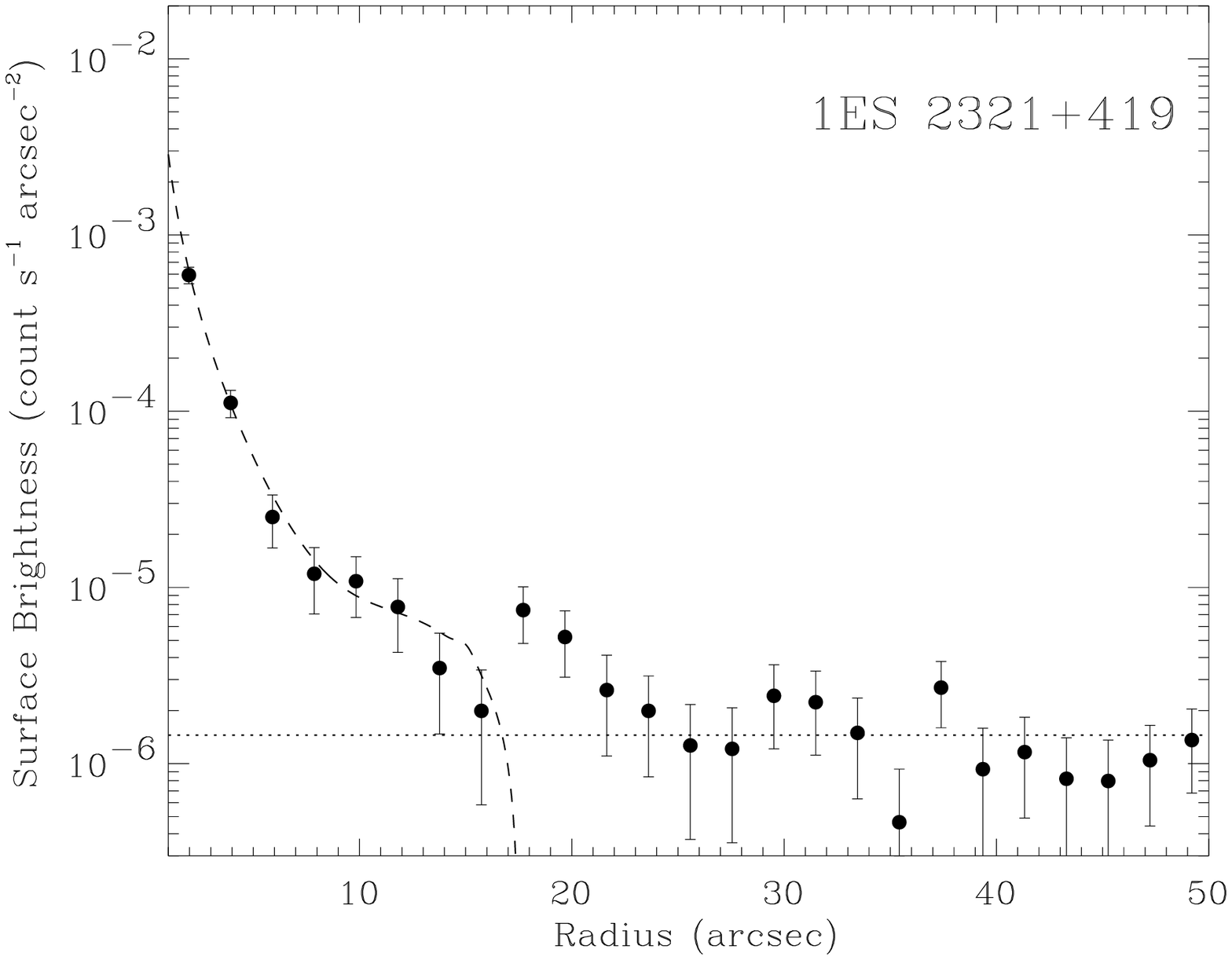}}{\includegraphics[width=9.1cm]{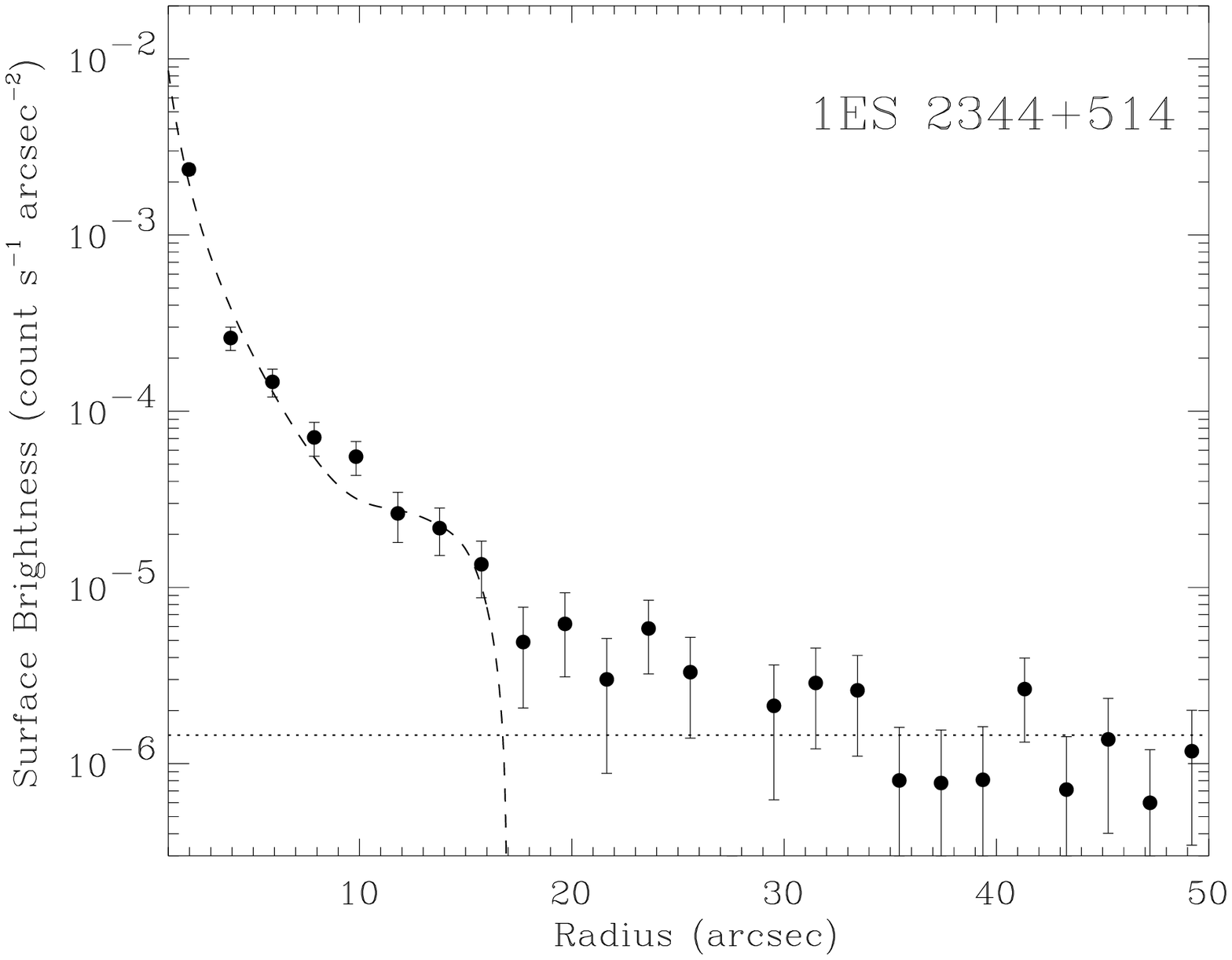}}
\caption[]{ACIS radial profiles for each source of the sample 
(data points). The composite PSF, derived by merging several
monochromatic PSFs (see text), is shown superimposed to the data
(dashed line); the horizontal dotted line marks the background. For
PKS 0548--322, PKS 2005--489, and BL Lac, excess X-ray counts over the
PSF are present at radii $\simgt$ 20\arcsec, indicating diffuse X-ray 
emission around the
core. Weak extended emission is also present for 1ES 2321+419 and 1ES
2344+514, while the profile for 3C~371 is entirely consistent with the
instrumental PSF.}
\end{figure*}
\clearpage

\noindent radial profiles with the PSF plus two $\beta$ models. In
the first model, the core radius was fixed to the halo radius of
23\arcsec\ and 12\arcsec\ for PKS 0548--322 and PKS 2005--489,
respectively. The parameters of the second $\beta$-model were left
free to vary. No improvement was obtained, and the core radius of the
second $\beta$ model is totally unconstrained. This is because the
core radius of the cluster extends on a scale larger than our field of
view, making its determination difficult. Thus, while the radial
profiles show clear evidence for the presence of X-ray emission on
cluster scales, it is not possible to constrain its spatial properties
with the present data. However, given its isotropic distribution, the
cluster X-ray emission dominates the X-ray spectrum, even when the
latter is extracted on the scale of the galaxy halo (see
below). Indeed, the cluster contribution is still present along the
line of sight when extracting the spectrum on small spatial scales.

To evaluate the luminosities of the diffuse emission, we extracted the
count rate from annular regions around the sources.  For BL Lac, the
annular region has inner and outer radii of 20\arcsec\and 35\arcsec,
respectively. The measured 0.3--8 keV count rate is 0.011$\pm$0.002
c/s. The count rate was converted into flux assuming a thermal
Raymond-Smith spectrum with Galactic column density, temperature
$kT=1$ keV, and fixing abundances to 0.2 solar.  The derived
observed flux is $F_{0.4-5\rm keV}=6.7\times 10^{-14}$ erg cm$^{-2}$
s$^{-1}$, corresponding to an absorption-corrected luminosity
$L_{0.4-5\rm keV}=1 \times 10^{42}$ erg s$^{-1}$, typical of FRIs
halos (e.g., Worrall 2002).

For PKS 0548--322 and PKS 2005--489, the integrated count rate in
20\arcsec--40\arcsec\ includes the contribution of the cluster. To
evaluate the halo luminosity, we subtracted the cluster contribution
from the total count rate in 20\arcsec\--40\arcsec, rescaling the
cluster surface brightness to the halo area and assuming a uniform
cluster emission profile. We derive 0.019$\pm$0.002 c/s and
0.038$\pm$0.003 c/s in 0.3--8 keV for PKS 0548--322 and PKS 2005--489,
respectively. Assuming a thermal Raymond-Smith spectrum with Galactic
column density, temperature $kT=1$ keV, and abundances fixed to
0.2 solar, these count rates correspond to an observed flux of
$F_{0.4-5\rm keV}=1.4 \times 10^{-13}$ erg cm$^{-2}$ s$^{-1}$ and
intrinsic luminosity $L_{0.4-5\rm keV}=1.3\times 10^{42}$ erg s$^{-1}$
for PKS 0548--322, and $F_{0.4-5\rm keV}=3.1\times 10^{-13}$ erg
cm$^{-2}$ s$^{-1}$ and $L_{0.4-5\rm keV}=3.1\times 10^{42}$ erg
s$^{-1}$ for PKS 2005--489.

An alternative way to estimate the luminosity of the halo is
based on the integration of the $\beta$-model over the extended
component region. Integrating between 20\arcsec\ and 40\arcsec\ we
obtained results fully consistent with those derived above.  With this
method, we can also evaluate the integrated luminosity over the entire
0\arcsec--40\arcsec\ range.  We found that the 0.4--5 keV unabsorbed
fluxes for BL~Lac, PKS 0548--322, and PKS 2005--489 are $F_{0.4-5\rm
keV}=2.8\times 10^{-13}$ erg cm$^{-2}$ s$^{-1}$, $F_{0.4-5\rm
keV}=3.3\times 10^{-13}$ erg cm$^{-2}$ s$^{-1}$, $F_{0.4-5\rm
keV}=7.6\times 10^{-13}$ erg cm$^{-2}$ s$^{-1}$, respectively.  The
corresponding intrinsic luminosities are $L_{0.4-5\rm keV}=2.9\times
10^{42}$ erg s$^{-1}$ for BL~Lac $L_{0.4-5\rm keV}=3.3\times 10^{42}$
erg s$^{-1}$ for PKS 0548--322 $L_{0.4-5\rm keV}=7.8\times 10^{42}$
erg s$^{-1}$ for PKS 2005--489.

The assumption of uniform cluster emission is unrealistic, as the
cluster radial profile may increase toward the center of the cluster,
where the gas is denser/hotter. Thus, the above count rates represent
an upper limit to the X-ray emission of the galaxy's halo.  For the
cluster emission, the count rates extracted from
40\arcsec--150/130\arcsec\ are 0.021 $\pm$ 0.002 c/s for
PKS~0548--322, and 0.013 $\pm$ 0.001 c/s for PKS~2005--489. Assuming
the parameters from the spectral analysis (see below), the
corresponding observed fluxes and intrinsic luminosities are
$F_{0.4-5\rm keV}=2.9\times 10^{-13}$ \flux\ and $L_{0.4-5\rm
keV}=3.\times 10^{42}$ \lum\ for PKS~0548--322, and $F_{0.4-5\rm
keV}=1.3\times 10^{-13}$ \flux\ and $L_{0.4-5\rm keV}=1.3\times
10^{42}$ \lum\ for PKS~2005--489, typical of cluster with Abell
richness 1 (Mulchaey \& Zabludoff 1998; Mushotzky 1998; Mahdavi et
al. 1997).

For 3C~371, 1ES 2321+419, and 1ES 2344+514, we evaluated upper limits
to the extended halo. Count rates were extracted in an annulus of
inner and outer radii 20\arcsec\ and 40\arcsec, respectively, and
converted into flux assuming a Raymond-Smith thermal model with $kT=1$
keV, abundances fixed to 0.2 solar, and Galactic column
densities. We derive $L_{0.4-5\rm keV}=1.2\times 10^{41}$ erg s$^{-1}$
for 3C 371, $L_{0.4-5\rm keV}=5.2\times 10^{40}$ erg s$^{-1}$ for 1ES
2321+419, and $L_{0.4-5\rm keV}=4 \times 10^{40}$ erg s$^{-1}$ for 1ES
2344+514.

In conclusion, diffuse X-ray emission is clearly present in 3 out of 6
sources of the sample (PKS 0548--322, PKS 2005--489, and BL Lac). In
two other sources, 1ES 2321+419 and 1ES 2344+514, there is some weak
evidence for faint diffuse emission. The diffuse emission in BL Lac,
and possibly in the two 1ES sources, is on the scale of the galaxy's
halo and has a luminosity typical of FRIs. In the two PKS sources,
there is evidence for diffuse emission on both the galaxy halo and the
cluster scale. However, it is not possible to model the cluster 
spatial properties.

\section{Spectral Analysis}

The goal of this section is to investigate the X-ray properties of the
diffuse X-ray emission in PKS 0548--322 and PKS 2005--489. To this
end, we extracted and analyzed the X-ray spectra of this component.

The ACIS-I spectra of the diffuse emission was extracted using an
annulus centered on the core. We used a large extraction region (inner
radius 20\arcsec, outer radius 150/130\arcsec) to increase the
signal-to-noise ratio. We chose this high value for the inner radius
in order to reduce contamination of the PSF wings from the AGN. An
issue with such a large extraction region, however, is that it
encompasses 4 CCDs that have different spectral responses. 
To circumvent this difficulty, we divided the extraction region 
in four
sectors, each one covering a single CCD. The resulting four spectra
were fitted jointly within XSPEC v.11.0.1.
To account for the recently observed quantum efficiency decay of ACIS,
possibly caused by molecular contamination of the ACIS filters, we
have applied a time-dependent correction to the ACIS quantum
efficiency based on the presently available information from the 
CXC\footnote[1]{http://cxc.har\-vard.edu/ciao/threads/ap\-ply\_acis\-abs/in\-dex.html}.

The model used to fit the spectra is the thermal plasma model APEC, 
with column density fixed to the Galactic value (Table 1), solar abundances 
fixed at 0.2, and temperature $kT$ ranging between 0 and 15 keV.  
We found $kT=6.3^{+8.7}_{-2.8}$ keV, $\chi^2_{\rm r}=0.42/47$, for PKS
0548-322, and $kT=4.6^{+2.0}_{-1.7}$ keV, $\chi^2_{\rm
red}=0.97/54$ for PKS 2005--489.  No improvements are obtained adding a
second component (a power law or a second thermal model). The fitted
temperatures are typical of the X-ray emission from the hot
intracluster gas in rich cluster (De Grandi \& Molendi 2002).

\begin{figure}
{\psfig{file=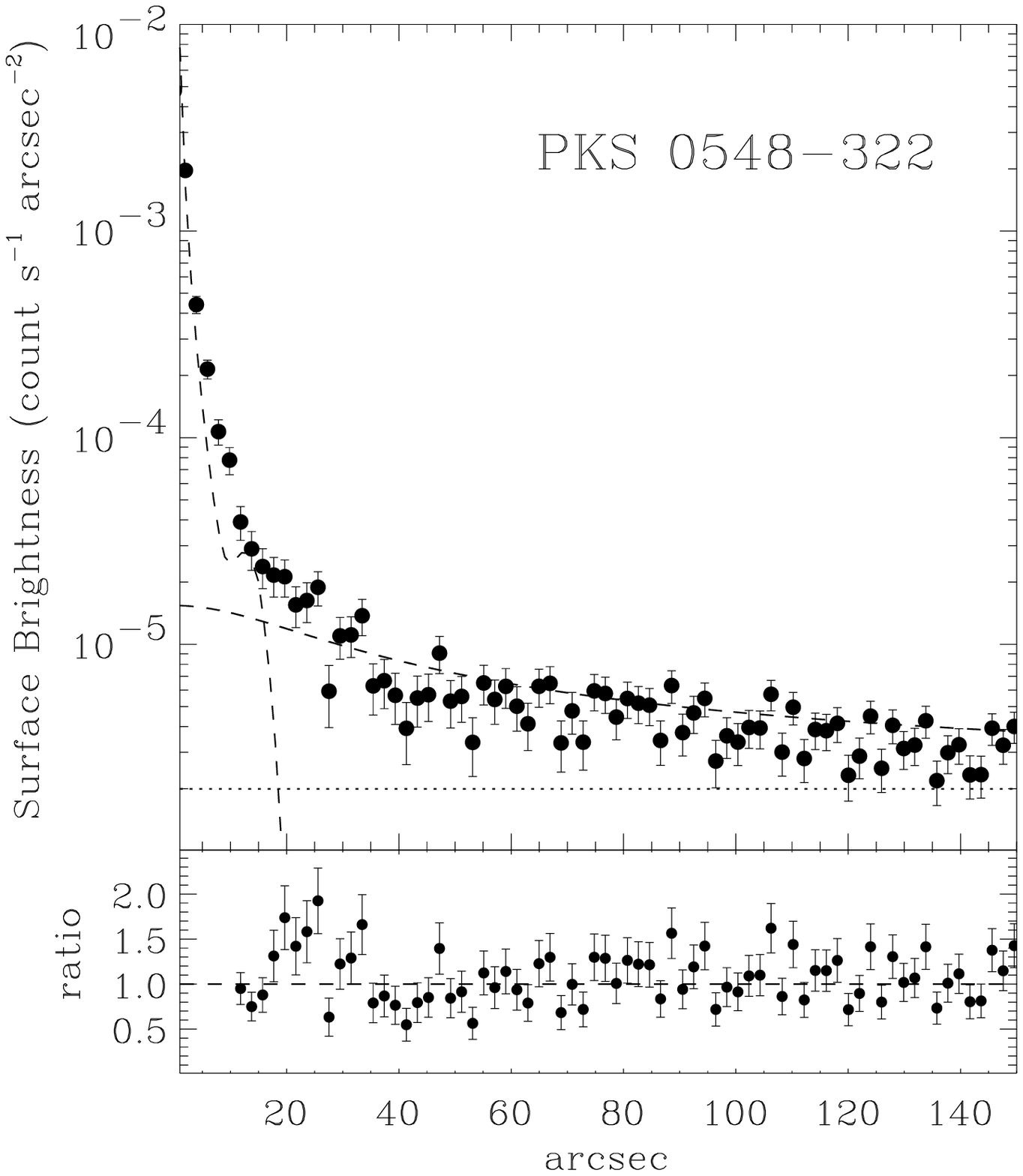,width=8.8cm,height=7.5cm,bbllx=55pt,bblly=29pt,bburx=489pt,bbury=520pt,clip=}}

{\psfig{file=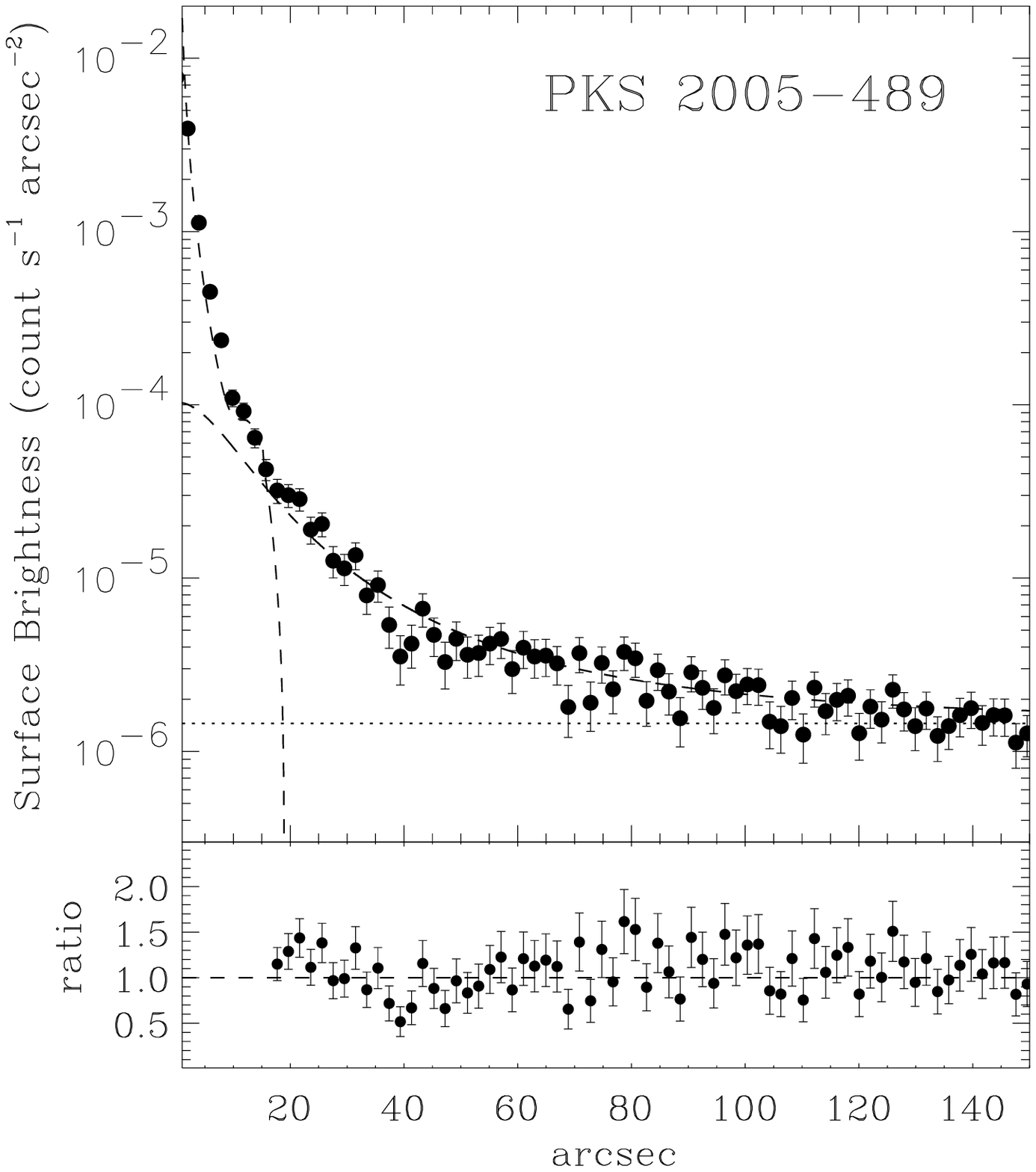,width=8.8cm,height=7.5cm,bbllx=55pt,bblly=29pt,bburx=489pt,bbury=520pt,clip=}}
\caption[]{Radial profiles and best-fit $\beta$-model for PKS 0548--322 and
PKS 2005--489. In both cases, a single $\beta$-model was fitted to the
data. However, two components for the diffuse emission are present,
the first one on the scale of the host galaxy ($\sim$ 40\arcsec), and
the second on the scale of the cluster ($\sim$ 130--150\arcsec). See
text for further details.}
\end{figure}

In an attempt to separate the contributions of the galaxy halo and of
the cluster, we extracted two X-ray spectra, the first in the range
20\arcsec--40\arcsec\ (to maximize the halo) and the second in
40\arcsec--130/150\arcsec, for both sources. We found that the best-fit
is obtained using the APEC model and abundances fixed to 0.2 solar. 
Spectral analysis of the first spectrum shows that the fitted temperature 
(with range between 0 and 15 keV) is still high ($kT=9.5^{+5.5}_{-4.6}$ keV 
with $\chi^2_{\rm r}=0.62/6$ for PKS 0548--322 and $kT=3.7^{+1.8}_{-1.3}$ keV 
with $\chi^2_{\rm r}=0.90/18$ for PKS 2005--489), indicating that the cluster 
emission is still dominant.
For the second spectrum, the parameters are: $kT=6.0^{+9.0}_{-2.9}$ keV, 
$\chi^2_{\rm r}=0.39/40$, for PKS 0548-322, and $kT=12.1^{+2.9}_{-8.1}$ 
keV, $\chi^2_{\rm r}=0.98/34$ for PKS 2005--489.

\section{Summary and Conclusions}

We presented \chandra\ ACIS-I/S observations of 6 BL Lacertae objects
(four HBLs and two LBLs), aimed at detecting the diffuse circumnuclear
X-ray emission predicted by the unification models. The short
exposures, 2--6 ks (10 ks for 3C 371), were optimized to detect
diffuse emission, but are insufficient to study their properties 
in detail.

Diffuse X-ray emission was convincingly detected in 3/6 cases (PKS
0548--322, PKS 2005--489, and BL Lac), with marginal evidence in an
additional 2 sources (1ES 2321+419 and 1ES 2344+514). The extended
X-ray emission is on scales of several kiloparsec and has a luminosity
similar to FRI galaxies studied with {\it Chandra} (Worrall et
al. 2002). This qualitatively supports the unification models for
radio-loud AGN, which states that BL Lacs and FRI galaxies are the
same intrinsic objects, seen at different orientations of the
relativistic jets, with BL Lacs being the sources more closely 
aligned with the line of sight.

The presence of an extended X-ray emission around the X-ray cores,
interpreted as a thermal bremsstrahlung from a galactic atmosphere,
can trace the medium whose pressure may confine the jets that are
pointing toward us.  For PKS 0548--322 and PKS 2005--489 we evaluated
the density and the external pressure of the X-ray component.  The gas
density profile is derived by deprojection of the $\beta$-model used
for the decomposition of the surface brightness profile (e.g., Ettori
2000). We adopted the cooling function value (e.g., Sarazin 1988) for
the hot gas temperature and abundance derived from the spectral
analysis.  The resulting central gas density is $1.4\times
10^{-3} ~\rm{atoms~cm}^{-3}$ for PKS 0548--322 and $7.4\times 10^{-3}
~\rm{atoms~cm}^{-3}$ for PKS 2005--489. The external pressures evaluated
at the characteristic radii (28 kpc for PKS 0548--322 and 15 kpc
for PKS 2005--489) are $P_{\rm ext} = 5.2\times 10^{-14} ~\rm{N~m}^{-2}$
and $P_{\rm ext} = 2.2\times 10^{-13} ~\rm{N~m}^{-2}$, respectively. 

Comparing the external gas pressure to the internal jet pressure
can provide useful information on the propagation of the jet. As the
jets of BL Lacertae objects are seen at small angles with respect to
the line of sight, the internal jet pressure can only be derived via
modeling of the SED of the blazar. Ghisellini et al. (2002) estimated
the total jet pressure for a sample of blazars, including FSRQs, LBLs,
and HBLs. In the case of our sources (considered extreme HBLs), 
the inferred pressure ranges from $\sim 2\times10^{-3}~\rm{N~m}^{-2}$ 
to $\sim 6\times10^{-2}~\rm{N~m}^{-2}$ (Fig. 2 of Ghisellini et al. 2002).  
These are several orders of magnitude larger than the
external gas pressure, implying the jets cannot be confined by the
external pressure of the galactic atmosphere, at least on parsec 
scales. This is not unexpected, since it is commonly accepted that the
jets start out supersonic near the central black hole.

One of the goals of the {\it Chandra} observations was to determine
whether HBLs and LBLs are characterized by different environments. The
two classes exhibit different properties in terms of their Spectral
Energy Distributions (e.g., Donato et al. 2001 and references
therein), and the decreasing luminosity from LBLs to HBLs can be
interpreted as a reduction in the jet power from the central energy
source. One can thus ask the question of whether the difference in jet
power in HBLs and LBLs is due to different gaseous environments. For
example, one could hypothesize that the less powerful HBL jets live in
a denser medium that is more efficient in decelerating
them. Unfortunately, the present sample is insufficient to draw firm
conclusions: of the 5 sources exhibiting diffuse X-ray emission, 4 are
HBLs and only 1 is an LBL. More fundamentally, the short ACIS exposure
prevents us from a detailed spectral analysis of the circumnuclear
gas. Deeper follow-up \chandra\ observations are needed to this aim,
as well as a larger, statistical sample of both HBLs and LBLs. On the
optical side, we note that recent \hst\ observations reveal no
differences in the host galaxy properties of the two BL Lac subclasses
(Urry et al. 2000, Scarpa et al. 2000).

Diffuse X-ray emission on scales of 100 kpc or more, typical of a
cluster of galaxies, is present in PKS 0548--322 and PKS
2005--489. This finding independently confirms previous claims, based
on optical imaging, that these two BL Lacs reside in clusters of
moderate-to-poor richness. However, the spatial parameters of the
X-ray emission of the cluster are unconstrained in our ACIS images,
while the X-ray emission of the cluster gas dominates the spectrum
integrated along the line of sight. The cluster temperature and
luminosity appears consistent (within 2$\sigma$) with the values found
for normal clusters, i.e., clusters with no AGN activity (e.g.,
Mahdavi et al. 1997, Mulchaey \& Zabludoff 1998; Mushotzky 1998;
Markevitch 1998), suggesting that the presence of the AGN does not
affect the global properties of the gas.  On the other hand, it is
also interesting that the two brightest X-ray BL Lacs of the sample
reside in an extended environment, suggesting that the gas can affect
nuclear activity. For example, Ellingson et al.  (1991)
suggested that nuclear activity could be triggered by galaxy-galaxy
interactions and merging in the cluster core. Such a possibility is
supported in the case of PKS 0548--322 by the presence of tidal
interactions between the host galaxy of the BL Lac and nearby
companion galaxies as seen in the optical image (Falomo et al. 1995).

In conclusion, we detected diffuse X-ray emission around the cores of
3, and possibly 5, BL Lacertae objects within our short \chandra\
exposures. The core radii and luminosities of the diffuse X-ray
emission are similar to those observed in FRI radio galaxies. This
finding supports current unification models for radio-loud AGN, which
attempt at unifying BL Lacs and FRI galaxies through
orientation. Future work will include deeper and additional X-ray
observations of a larger, statistical sample of BL Lacs, in order to
put this conclusion on a firmer basis and to determine in greater
detail the physical properties of the diffuse gas.

\acknowledgements{
We thank the referee, Dr. M.J. Hardcastle, for the useful comments and
suggestions that improved the paper.
We gratefully acknowledge financial support from NASA grants
NAG5-10073 (DD, RMS), NAS8-39073 (JEP), and LTSA grant NAG5-10708 (MG,
RMS).
This research made use of the NASA/IPAC Extragalactic Database (NED)
which is operated by the Jet Propulsion Laboratory, Caltech, under
contract with the National Aeronautics and Space Administration.}

\begin{center}
{\bf APPENDIX A

Serendipitous sources in the \chandra\ fields}
\end{center}

A glance at the \chandra\ images reveals the presence of several point
sources in the fields of the targets. We used the \verb+CIAO+ tool
{\tt wavdetect} to search for serendipitous X-ray sources in the
f.o.v. In the algorithm, the parameter {\tt scale} (a list of radii,
in image pixels, of Mexican Hat wavelet functions) was left free to
range between 1 and 16, and the parameter {\tt threshold} (the number
of detected spurious sources in a pixel map) held fixed at
{$10^{-9}$}.  The algorithm returns a list of elliptical regions that
define the positions and the shapes of the detected sources. We next
used the coordinates from {\tt wavdetect} and its associated error
regions to search for their optical counterparts on ESO archival
plates. We used circular search regions with radii of 5\arcsec\
because it is known that most reprocessed ACIS-I observations have an
offset of up to 1.5\arcsec. \footnote[2]{See the memo on astrometry
problems at http://cxc.\
harvard.edu/\-mta/\-ASPECT/\-improve\_as\-trom\-e\-try.html}

In Table 5 we list the coordinates (J2000), the net counts, and the
detection significance in $\sigma$, where $\sigma$ is the ratio of the
net source counts to the ``Gehrels error'' of the background counts
(for more details, see {\tt wavdetect} manual at the following
URL http://asc.harvard.edu/toolkit/pimms.jsp). 
We report
the hardness ratios, defined as the difference between the count rates
in the 2--8 keV (hard band) and those in the 0.3--2 keV (soft band),
divided by the 0.3--8 keV count rate. 
A positive value of the hardness
ratio suggests that the source has a flat spectrum, while a positive
value suggests a steep spectrum.  Also listed in Table 5 are the ESO
identification codes, the angular distance of the optical source from
the X-ray source, and the apparent magnitudes in the red and blue
bands. The NED and Simbad on-line catalogs were searched to find
additional information on the optical counterparts.

We now comment on individual \chandra\ fields.

\begin{figure}
{\psfig{file=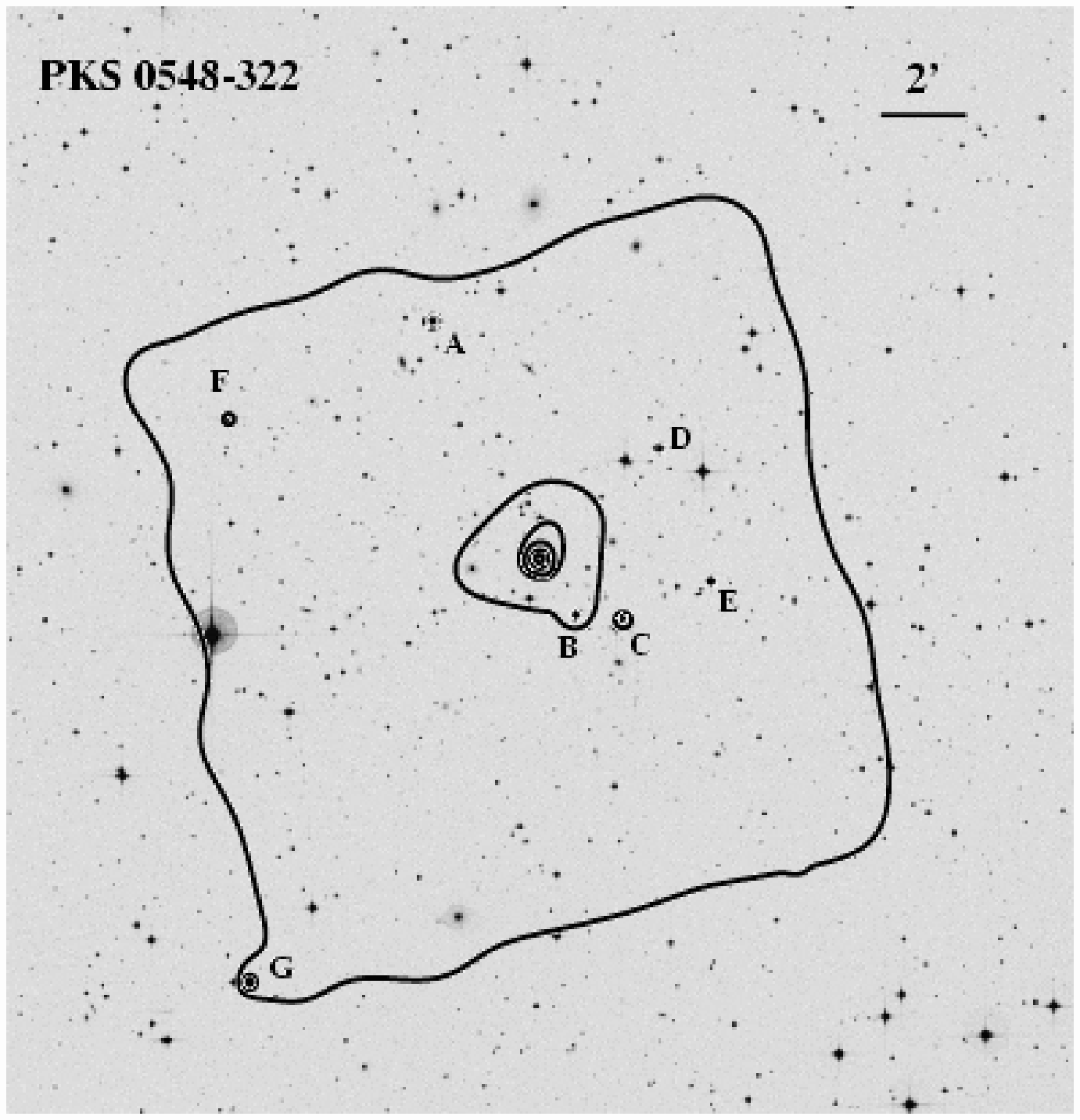,width=9.cm}}
{\psfig{file=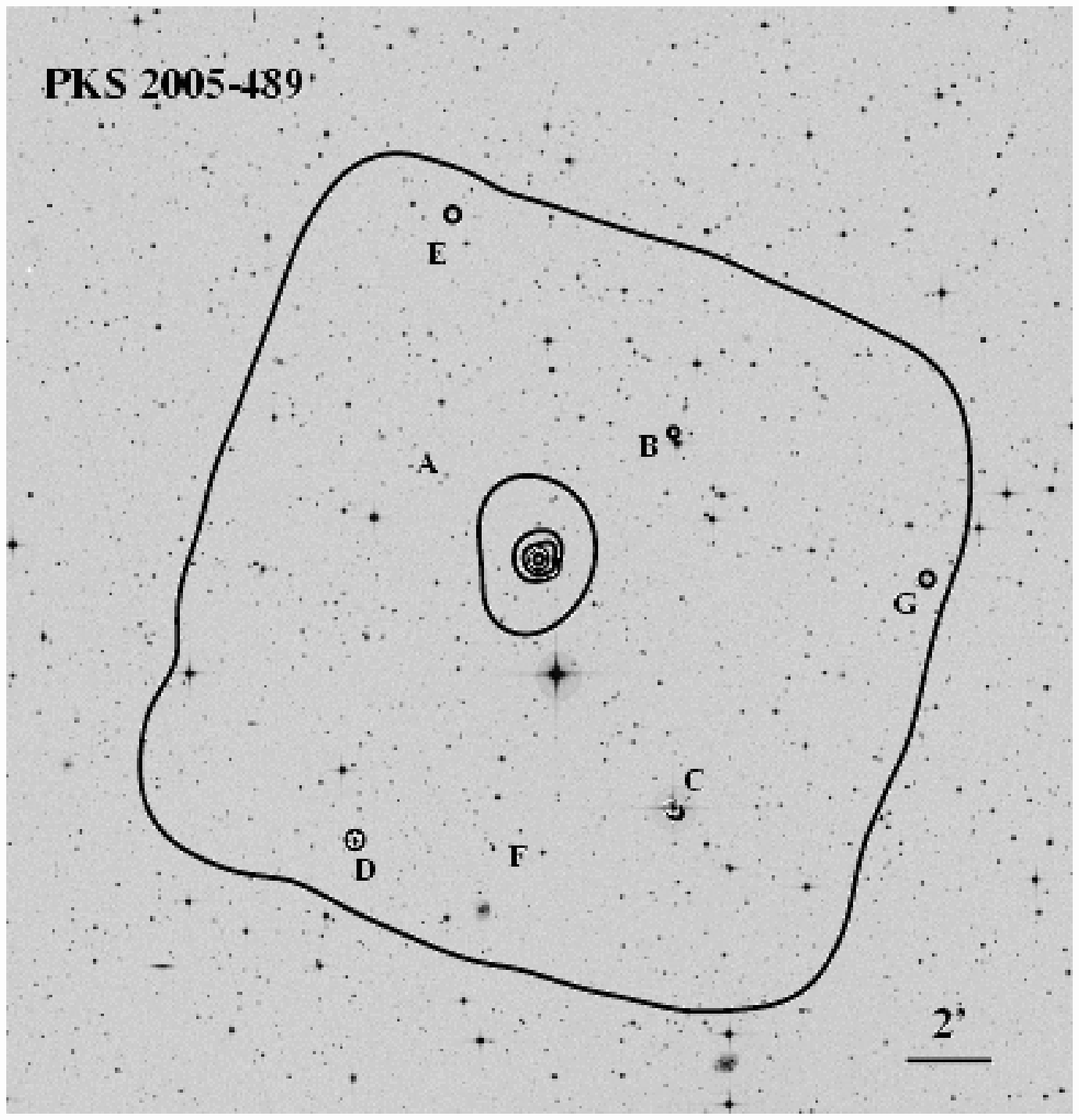,width=9.cm}}
\caption[]{X-ray contours in the energy range 0.3--8 keV overlaid 
on the optical images obtained from the 2.4 m Mount Palomar Telescope
(POSS) archive for PKS 0548--322 (top) and PKS 2005--489 (bottom). The
serendipitous X-ray sources detected with {\tt wavdetect}
(see text and Table 5) are marked with capital letters.
North is to the top and East to the left.}
\end{figure}

\clearpage
\begin{table*}
\caption{Field Sources}
\begin{center}
\begin{tabular}{clccccccccc}
\hline
\noalign{\smallskip}
   &~~~~ R.A.    & Dec.       & $c/s$ & HR  & ($\sigma$)& ESO Code     & Dist.   &$M_R$&$M_B$&Note\\
(1)&~~~~~ (2)     &  (3)       &  (4)  & (5)   &   (6)     &  (7)         & (8)     & (9)  & (10)  &(11)\\
\noalign{\smallskip}
\hline
\noalign{\smallskip}
&PKS 0548--322 &&&&&&&&&\\
A & 05 50 52.17 &  -32 10 52.5  &  91 & -0.65 &   36.85 & U0525-02418996 &  0.560  & 16.3 &  16.8 &\\
B & 05 50 36.63 &  -32 17 33.3  & 193 & -0.59 &   74.42 & U0525-02415715 &  0.406  & 14.6 &  14.8 &(a)\\
C & 05 50 31.50 &  -32 17 39.2  &  28 & -0.86 &   11.49 & U0525-02414666 &  0.506  & 17.8 &  20.1 &\\
D & 05 50 27.25 &  -32 13 49.0  &  17 & -0.65 &    7.62 & U0525-02413729 &  0.563  & 18.1 &  20.7 &\\
E & 05 50 21.63 &  -32 16 49.4  &  18 & -0.22 &    7.97 & U0525-02412549 &  0.727  & 17.9 &  18.9 &(b)\\
F & 05 51 14.78 &  -32 13 03.4  &  12 & -0.83 &    5.12 & U0525-02423572 &  1.284  & 11.9 &  12.0 & \\
G & 05 51 13.26 &  -32 25 47.5  &  32 &  0.06 &   11.11 &              - &    -    &     -&    -  & \\
\hline
\noalign{\smallskip}
&3C 371	&&&&&&&&&\\			  
A & 18 07 37.87 &  69 51 12.1   &   8 &  0.00 &    3.61 & -              &    -   &    - &    - &\\
B & 18 07 26.57 &  69 46 26.6   &  38 & -0.79 &   11.26 & U1575-03911444 & 0.521  & 19.2 & 19.2 &\\
C & 18 07 04.30 &  69 49 32.8   &   7 &  0.14 &    2.88 & -              &    -   &    - &    - &\\
D & 18 06 50.17 &  69 46 23.1   &   7 & -1.00 &    2.59 & -              &    -   &    - &    - &\\
E & 18 07 27.11 &  69 45 32.1   &  28 & -0.79 &    7.87 & -              &    -   &    - &    - &\\
F & 18 06 58.78 &  69 43 59.0   &  36 & -0.17 &   10.59 & -              &    -   &    - &    - &\\
G & 18 06 03.36 &  69 51 17.4   &  19 & -0.79 &    7.34 & U1575-03907608 & 1.382  & 16.9 & 19.3 &\\
\hline
\noalign{\smallskip}
&PKS 2005--489 &&&&&&&&&\\			  
A & 20 09 41.06 & -48 47 22.9  &   8 & -0.25 &  3.76  &  -              &    -   &    -   &    -   &\\
B & 20 09 13.88 & -48 46 38.0  &   8 & -0.25 &  3.74  &  -              &    -   &    -   &    -   &\\
C & 20 09 07.14 & -48 55 35.2  &  33 & -1.00 & 14.55  & U0375-38531345  & 0.385  &   9.7  &  11.4  &(c)\\
D & 20 09 52.53 & -48 56 08.9  &  30 & -0.20 & 12.43  & U0375-38543383  & 0.536  &  18.2  &  19.0  &\\
E & 20 09 36.39 & -48 42 02.9  &  11 & -0.27 &  4.59  & U0375-38539109  & 1.104  &  18.2  &  19.4  &\\
F & 20 09 30.34 & -48 56 19.8  &   7 & -0.71 &  3.53  & U0375-38537485  & 1.560  &  18.2  &  19.9  &\\
G & 20 08 30.15 & -48 50 33.2  &  17 & -0.53 &  6.06  & -               &    -   &    -   &    -   &\\
\hline
\noalign{\smallskip}
&BL LAC	&&&&&&&&&\\			  
A & 22 02 53.68 &    42 17 48.5 &  32 & -0.94 &  16.11 &  U1275-16943177 &  2.131 &   13.0 &   14.5 &\\
B & 22 02 51.03 &    42 13 02.2 &  15 & -0.07 &   7.28 &  U1275-16942082 &  0.588 &   18.0 &   18.2 &\\
C & 22 02 43.81 &    42 13 45.3 &   9 &  0.56 &   4.90 &   -             &      - &    -   &    -   &\\
D & 22 02 14.64 &    42 13 53.4 &  38 & -0.58 &  18.69 &   -             &      - &    -   &    -   &\\
E & 22 03 11.04 &    42 14 36.9 &  10 & -0.20 &   4.83 &   -             &      - &    -   &    -   &\\
F & 22 03 09.86 &    42 13 37.0 &   6 & -0.33 &   3.40 &  U1275-16950337 &  0.420 &   13.9 &   15.0 &\\
G & 22 03 03.50 &    42 14 01.2 &   8 & -0.50 &   4.22 &   -             &      - &    -   &    -   &\\
H & 22 02 54.06 &    42 11 03.6 &  19 & -0.26 &   8.94 &  U1275-16943366 &  2.604 &   17.4 &   18.1 &\\
I & 22 02 30.81 &    42 11 29.2 &   7 &  0.14 &   3.32 &   -             &      - &    -   &    -   &\\
L & 22 02 18.88 &    42 22 28.9 &  41 &  0.07 &   1.88 &   -             &      - &    -   &    -   &\\
\hline
\noalign{\smallskip}
&1ES 2321+514	&&&&&&&&&\\		  
A & 23 24 06.48 &   42 11 52.9  & 16 &  0.12 &   8.27 &   -             &      -   &    -   &    -   &\\
B & 23 23 40.64 &   42 05 29.0  & 13 &  0.38 &   6.34 &  U1275-18369935 &  0.463   &   17.9 &   18.6 &\\
C & 23 23 35.36 &   42 05 51.8  & 18 & -0.78 &   8.70 &   -             &      -   &    -   &    -   &\\
D & 23 23 33.08 &   42 07 42.1  &  9 & -0.33 &   4.18 &   -             &      -   &    -   &    -   &\\
E & 23 24 16.16 &   42 14 28.7  & 12 & -0.33 &   5.94 &   -             &      -   &    -   &    -   &\\
F & 23 24 02.15 &   42 05 21.2  &  8 &  0.50 &   4.25 &  U1275-18374011 &  1.440   &   16.4 &   17.1 &\\
G & 23 23 27.33 &   42 08 45.9  &  8 &  0.25 &   3.57 &  -              &     -    &    -   &    -   &\\
H & 23 24 01.26 &   42 00 46.5  & 18 & -0.56 &   6.13 &  -              &     -    &    -   &    -   &\\
I & 23 23 43.62 &   42 01 44.4  & 13 & -0.54 &   5.22 &  U1275-18370478 &  3.179   &   18.4 &   18.6 &\\
L & 23 23 26.60 &   42 05 53.7  & 13 & -0.54 &   5.51 &   -             &      -   &    -   &    -   &\\
M & 23 23 11.51 &   42 06 36.6  & 17 & -0.53 &   5.69 &   -             &      -   &    -   &    -   &\\
\hline
\noalign{\smallskip}
&1ES 2344+514	&&&&&&&&&\\		  
A & 23 47 27.14 &   51 40 18.4 &   9 & -0.56 &   5.56 &  U1350-18688025  &  1.522  &  18.7  &  19.1 &\\
B & 23 47 10.40 &   51 40 32.3 &   8 & -0.25 &   3.90 &  U1350-18682932  &  0.944  &  16.5  &  17.9 &\\
C & 23 46 36.64 &   51 42 30.5 &  16 & -0.50 &   7.86 &   -              &      -  &    -   &   -   &\\
D & 23 47 34.55 &   51 43 52.2 &   7 & -1.00 &   3.36 &  U1350-18690283  &  0.840  &  11.2  &  11.8 &(d)\\
E & 23 47 23.94 &   51 41 36.8 &   7 & -0.43 &   3.47 &   -              &      -  &    -   &   -   &\\
F & 23 48 06.02 &   51 40 16.9 &   9 & -0.56 &   4.18 &   -              &      -  &    -   &   -   &\\
G & 23 47 32.77 &   51 34 42.0 &   4 & -1.00 &   1.81 &   -              &      -  &    -   &   -   &\\
\hline
\end{tabular}
\end{center}
{\bf Columns}: 1=Detected source; 2=Right
Ascension at J2000; 3=Declination at J2000; 4=Net count rate in 0.3--8
keV; 5=Hardness ratio, defined as $(h-s)/(h+s)$ where $h$ is the
count rate in 2--8 keV and $s$ in 0.3--2 keV; 6=Detection significance
(in $\sigma$, see Sect. 7); 7=ESO identification code; 8=Distance (in 
\arcsec) between the Chandra coordinate and the optical one; 
9-10=Optical magnitude 
in the red and blue band; 11=(a) Galaxy 2MASXi J0550366-321733; (b)
ROSAT X--ray source 1WGA J0550.3-3216 XrayS; (c) Star CCDM
J20091-4856AB or HD 190857; (d) Star GSC 03650-00158.
\end{table*}
\clearpage

\noindent 

{\bf PKS 0548--322}: In the {\it Chandra} field we found
7 serendipitous point sources (Table 5).  Optical counterparts are
found for all sources, except source G (see Fig. 5, left, for the
overlay of the X-ray isocontours on the optical image). The X--ray
sources detected with {\tt wavdetect} are marked with capital
letters. Source E coincides with 1WGA J0550.3-3216, an X--ray source
previously identified by the \rosat\ satellite.  Source B is a galaxy
from the 2MASS catalog (2MASXi J0550366-321733), studied in previous
optical works (Pesce et al. 1995, Falomo et al.  1995). This spiral
galaxy coincides with galaxy G4 in the PKS 0548--322 cluster (Falomo et
al. 1995), and has a redshift of $z$=0.072. Falomo et al. (1995) found
strong emission lines in the optical spectrum and classified the
source as an extreme Fe II emitting AGN.  We measure $\sim$200 X-ray
counts from the nucleus of this galaxy in the energy range 0.3--8 keV,
sufficient for a crude spectral analysis. The X-ray spectrum is well
fitted ($\chi^2_{\rm r}=1.37/16$) by a single power law with
absorption fixed to Galactic, and photon index $\Gamma=2.21\pm
0.33$. This is consistent with the canonical slope of $\Gamma$=1.8
measured for Seyferts and other lower-luminosity AGN. The observed
2--10 keV flux is $F_{2-10 \rm keV}=1.4\times 10^{-13}$ ergs
cm$^{-2}$ s$^{-1}$, corresponding to an intrinsic luminosity of
$L_{2-10 \rm keV}=1.4\times 10^{42}$ erg s$^{-1}$, assuming a
redshift of 0.072.

{\bf 3C~371}: The Western elongation of the soft X-ray emission in Fig. 1 is
due to the X--ray jet (Pesce et al. 2001).  Seven X--ray sources are
detected in the {\it Chandra} field but only two sources, B and G,
have an optical counterpart.

{\bf PKS 2005--489}: In the X--ray image we detected 7 point sources.  
Sources D, E, and F have very weak optical counterparts,
and source C corresponds to a F8V double or multiple star (CCDM
J20091-4856AB also called HD 190857). For the positions of the
detected X--ray sources on the optical image, see Fig. 5, right.

{\bf BL~Lac}: Ten serendipitous sources are detected in the {\it
Chandra} field. Only four of them (A, B, F, and H) have optical
counterparts. 

{\bf 1ES 2321+419}: Eleven field sources are detected.  Only three of
them (B, F, and I) have optical counterparts. 

{\bf 1ES 2344+514}: Seven X-ray sources are found in this \chandra\ field. 
The weak X--ray source D coincides with a bright star (GSC 03650-00158).

\end{document}